\documentclass[11pt,a4paper]{article}
\usepackage{amsfonts}
\usepackage{amsmath}
\usepackage{amssymb}
\usepackage{graphicx}
\usepackage{geometry}
\usepackage{cancel}
\usepackage{hyperref}
\newtheorem{theorem}{Theorem}

\newtheorem{corollary}{Corollary}

\newtheorem{remark}{Remark}

\newenvironment{proof}[1][Proof]{\textbf{#1.} }{\  \rule{0.5em}{0.5em}}
\makeatletter
\def \@removefromreset#1#2{\let \@tempb \@elt
\def \@tempa#1{@&#1}\expandafter \let \csname @*#1*\endcsname \@tempa
\def \@elt##1{\expandafter \ifx \csname @*##1*\endcsname \@tempa \else
\noexpand \@elt{##1}\fi}     \expandafter \edef \csname cl@#2\endcsname{\csname cl@#2\endcsname}     \let \@elt \@tempb
\expandafter \let \csname @*#1*\endcsname \@undefined}

\@removefromreset{equation}{section}

\@removefromreset{theorem}{section}
\makeatother

\begin{document}
\title{High-spin measurements in an arbitrary two-qudit state}
\author{Elena R. Loubenets$^{1,2}$ and Louis Hanotel$^{1}$ 
\\ 
$^{1}$ Department of Applied Mathematics, MIEM, HSE University, \\
Moscow 123458, Russia\\
$^{2}$Steklov Mathematical Institute of Russian Academy of Sciences, \\
Moscow 119991, Russia}
\date{}
\maketitle

\begin{abstract}
Violation of the CHSH inequality by a bipartite quantum state is now used in
many quantum applications. However, the explicit analytical expression for the maximal value of the CHSH expectation under 
local Alice and Bob spin-$s$ measurements is still known only for $s=1/2$. In the present article, 
for an arbitrary state of two spin-$s$ qudits, each of dimension $d=2s+1\geq 2$, we introduce the notion of the spin-$s$ correlation
matrix, which has dimension $3\times 3$ for all $s\geq \frac{1}{2}$; establish its relation to the general correlation $(d^{2}-1)\times (d^{2}-1)$ matrix of this state within the generalized Pauli representation and derive in
terms of the spin-$s$ correlation
matrix the explicit analytical expression for the maximal value of the CHSH expectation under local Alice and Bob spin-$s$ measurements in this state. 
Specifying this general expression for the two-qudit GHZ state,
the nonlocal two-qudit Werner state, and some nonseparable pure two-qudit states, we find that, under local Alice and Bob high-spin ($s\geq1$) measurements in each of these nonseparable states, including the maximally entangled one, the CHSH inequality is
not violated. Moreover, unlike the case of spin-$1/2$ measurements, where each pure nonseparable two-qubit state violates the CHSH inequality and the maximal  value of its CHSH expectation increases monotonically with a growth of its entanglement, the situation under high-spin measurements is quite
different -- for a pure two-qudit state with a higher degree of
entanglement, the maximal value of the CHSH
expectation turns out to be less than for a pure two-qudit state with lower entanglement and even for a separable one.

\end{abstract}
\maketitle
\section{Introduction}

Since the seminal paper of Bell \cite{Bel:64}, violation by bipartite quantum
states of Bell inequalities\footnote{
For the general framework on multipartite Bell inequalities with an arbitrary number of
settings and any type of outcomes at each of sites, see \cite{Lou:08}.} has been intensively discussed in the literature and is now used in many
quantum applications, see \cite{Bru.etal:14, Lou:17} and references therein.

The Clauser--Horne--Shimony--Holt (CHSH) inequality \cite{Cla.Hor.Shi:69}, which is
one of the mostly studied and used in quantum applications, concerns a value of the
quantum expectation 
\begin{align}
& \mathrm{tr}[\rho \text{ }\mathcal{B}%
_{chsh}(A_{1},A_{2};B_{1},B_{2})],  \label{0_0} \\
\mathcal{B}_{chsh}(A_{1},A_{2};B_{1},B_{2}) :&=A_{1}\otimes
B_{1}+A_{1}\otimes B_{2} +A_{2}\otimes B_{1}-A_{2}\otimes B_{2},  \notag
\end{align}
in a bipartite quantum state $\rho $ on $\mathcal{H}_{1}\otimes \mathcal{H}%
_{2}\mathbb{\ }$ under a correlation scenario where two participants, say Alice and Bob, measure observables $A_{i}$, $i=1,2,$ and $%
B_{j},j=1,2,$ on Hilbert spaces $\mathcal{H}_{1}$ and $\mathcal{H}_{2},$ respectively.

If a bipartite state $\rho $ admits a local hidden variable (LHV) model and quantum observables $A_{i}$, $B_{j}$ have  eigenvalues in $[-1,1]$,
then\footnote{Within the original derivation of the CHSH inequality in \cite{Cla.Hor.Shi:69}, each of quantum observables in the
left-hand side of (\ref{0_33}) has only two eigenvalues $\pm 1.$} 
\begin{equation}
\left\vert \text{\ }\mathrm{tr}[\rho \text{ }\mathcal{B}%
_{chsh}(A_{1},A_{2};B_{1},B_{2})]\text{ }\right\vert _{_{LHV}}\leq 2,  \label{0_33}
\end{equation}%
and specifically this inequality, holding in an LHV case, is referred to  as the CHSH one in the quantum information literature. 

As shown by Tsirelson \cite{7,8}, for any bipartite quantum state and any quantum observables $A_{i}, B_{j}, i,j = 1, 2,$ with eigenvalues in $[-1,1]$, the absolute value of the quantum CHSH expectation (\ref{0_0}) is upper bounded by $2\sqrt{2}$. 
This quantum upper bound is attained at each of the two-qubit Bell states, more generally, at each \cite{1, Lou:20} of the maximally entangled pure states of two qudits of an even dimension $d\geq2$, in particular, at the two-qudit Greenberger--Horne--Zeilinger (GHZ) state and at \cite{new1} the two-qudit singlet state. 

It was also proved in \cite{Lou:20} that, for a maximally entangled pure two-qudit state of an odd dimension $d\geq 3$, the maximum of the absolute value of the CHSH expectation (\ref{0_0}) over all traceless observables with eigenvalues in $[-1,1]$ is equal\footnote{This exact result \cite{Lou:20} for an odd $d\geq 3$  answers negatively on the question, posed by Gisin and Peres in \cite{1} -- whether for an odd $d\geq 3$ the Tsirelson upper bound could be attained at a set of observables different from those specified in \cite{1}.}  to $\frac{2(d-1)}{d}\sqrt{2}$. 

For $d=2$, the explicit analytical expression for the maximum $\Upsilon_{chsh}^{(traceless)}(\rho_{2\times 2})$ of the absolute value of the CHSH expectation (\ref{0_0}) in a state $\rho_{2\times 2}$ (pure or mixed) over all traceless qubit observables $A_{i}, B_{j}, i,j = 1, 2,$ was found in 
 \cite{Hor.Hor.Hor:95} and reads
\begin{equation}
\Upsilon_{chsh}^{(traceless)}(\rho_{2\times 2}) =2\sqrt{\tau _{1}^{2}(\rho _{2\times 2})+\tau _{2}^{2}(\rho _{2\times 2})}\ ,
\label{0_4}
\end{equation}%
where $\tau _{1,2}(\rho _{2\times 2})$  are
two largest singular values of the $3\times 3$ matrix $T_{d=2}\left( \rho
_{2\times 2}\right)$, referred to as the correlation matrix of  a  state $\rho _{2\times 2}$
and defined by
\begin{equation}
T_{d=2}^{(ij)}\left( \rho _{2\times 2}\right) :=\mathrm{tr}\left[ \rho
_{2\times 2}(\sigma _{i}\otimes \sigma _{j})\right] ,\ \ i,j=1,2,3.
\label{0_5}
\end{equation}%
Here, $\sigma _{j},\ j=1,2,3,$ are the Pauli operators on $\mathbb{C}^{2}$. 
By Eqs. (\ref{0_33}) and (\ref{0_4}) a two-qubit state $\rho _{2\times 2}$, pure or mixed, violates the CHSH inequality under Alice and Bob measurements of traceless qubit observables iff the Horodecki parameter \cite{Hor.Hor.Hor:95}
\begin{equation}
M_{hor}(\rho _{2\times 2}):=\sqrt{\tau _{1}^{2}(\rho _{2\times 2})+\tau
_{2}^{2}(\rho _{2\times 2})}>1.  \label{---7}
\end{equation}%
It was further shown\footnote{Relation (\ref{__8}) was first proved in \cite{Ver.Wol:02} for a pure two-qubit state of the specific form  and further in \cite{ Lou.Kuz.Han:24} for an arbitrary pure two-qubit state, see Remark 1 in \cite{Lou.Kuz.Han:24}.} in \cite{Ver.Wol:02, Lou.Kuz.Han:24} that, for 
every pure two-qubit state $|\psi _{2\times 2}\rangle \langle \psi _{2\times 2}|$,  
\begin{equation}
M_{hor}(|\psi _{2\times 2}\rangle )=\sqrt{1+\mathrm{C}^{2}(|\psi _{2\times
2}\rangle )}\ ,  \label{__8}
\end{equation}%
where $\mathrm{C}(|\psi _{2\times 2}\rangle )$ is the concurrence \cite%
{12, 12.1} of a pure two-qubit state $|\psi _{2\times 2}\rangle $.

For the maximum $\Upsilon_{chsh}^{(traceless)}(\rho_{d\times d})$ of the absolute value of the CHSH expectation (\ref{0_0}) over all traceless observables with eigenvalues in $[-1,1]$ in case of an arbitrary $d>2$,  the explicit analytical expression in terms of a two-qudit state $\rho_{d\times d}$ is not known. 

However, the general upper and lower bounds on $\Upsilon_{chsh}^{(traceless)}(\rho_{d\times d})$ were found in \cite{Lou:20, Lou.Kuz.Han:24} and, for a pure two-qudit state, the tight lower bound in \cite{Lou.Kuz.Han:24} reads\footnote{See Eqs. (39) and (47) in \cite{Lou.Kuz.Han:24}.}
\begin{equation}
\Upsilon_{chsh}^{(traceless)}(|\psi_{d\times d}\rangle)\geq2\sqrt{1+\frac
{1}{(2d-3)^{2}}\mathrm{C}^{2}(|\psi_{d\times d}\rangle)}\ , \label{0_6}%
\end{equation}
where $\mathrm{C}(|\psi _{d\times d}\rangle )$ is the concurrence \cite%
{13} of a pure two-qudit state $|\psi _{d\times d}\rangle
$ and the equality holds in a two-qubit case. 

This lower bound explicitly indicates that every pure nonseparable two-qudit state violates the CHSH inequality -- the well-known result, first proved in \cite{1} via presenting for every pure two-qudit state the specific dichotomic qudit observables for which the CHSH inequality is violated. 

Recall that, in a two-qubit case, every traceless qubit observable has, up to a real coefficient, the form $\sigma_{n}:=n\cdot\sigma=\sum_{k}n_{k}\sigma_{k}$, where $\sigma:=(\sigma_{1},\sigma_{2},\sigma_{3})$ and $n$ is a unit
vector in $\mathbb{R}^{3}$. Every spin-$\frac{1}{2}$ observable of a spin qubit is given by $\frac{1}%
{2}\sigma_{n}$. 
Therefore, for an arbitrary  state $\rho_{2\times 2}$ of two spin $s=\frac {1}{2}$ qubits, relation (\ref{---7}) constitutes also the necessary and sufficient condition for violation of the CHSH inequality under local Alice and Bob measurements of spin-$\frac {1}{2}$
observables, and, for the absolute value of the CHSH
expectation (\ref{0_0}) in this state, the maximum $\Upsilon_{chsh}^{(spin-s)}(\rho_{2\times 2})$ over all local Alice and Bob spin-$\frac{1}{2}$ observables is related to  the maximum (\ref{0_4}) over  all traceless qubit observables with eigenvalues in $[-1,1]$ as 
\begin{equation}
\Upsilon_{chsh}^{(spin-s)}(\rho_{2\times 2}) =\frac{1}{4}\Upsilon_{chsh}^{(traceless)}(\rho_{2\times 2}).
\end{equation}

However, in case of a spin  $s\geq 1$ qudit of dimension $d=2s+1$, while any  spin-$s$ observable is traceless, an arbitrary traceless qudit observable does not need to be proportional to some spin-$s$ observable, so that all spin-$s$ observables\footnote{Eigenvalues of a spin-$s$ observable are in $[-s,s]$.}, when normalized to have their eigenvalues in $[-1,1]$, are included into the set of all traceless observables with eigenvalues in $[-1,1]$ only as a particular subset. Therefore, for spin $s\geq 1$, the relation between two maximums -- over all traceless observables with eigenvalues in $[-1,1]$ and all spin-$s$ observables with eigenvalues in $[-s,s]$, takes the form: 
\begin{equation}
\frac{1}{s^{2}} \Upsilon_{chsh}^{(spin-s)}(\rho_{d\times d}) \leq \Upsilon_{chsh}^{(traceless)}(\rho_{d\times d}).\label{_9}
\end{equation}

From relations (\ref{0_6}) and (\ref{_9}) it follows that, though every pure nonseparable state of two spin qudits $d=2s+1, s\geq 1$ violates the CHSH inequality for some traceless qudit observables ${A_{j}, B_{k}, j,k=1,2}$ in (\ref{0_0}) \emph{this state does not need to violate the CHSH inequality for some spin-$s$ observables}. 

Note that the specific qudit observables with only two eigenvalues $\pm 1$, chosen in \cite{1} to show that every pure nonseparable state of two spin $s\geq 1$ qudits of dimension $d=2s+1$ violates the CHSH inequality, \emph{do not constitute spin $s\geq 1$ observables} -- either by the number of their eigenvalues or by their form. Recall that each spin-$s$ observable has $(2s+1)$ eigenvalues and cannot be dichotomic for $s\geq 1$.

Though the CHSH inequality is one of the mostly used in quantum applications, for an arbitrary state of two spin-$s$ qudits, the explicit analytical expression for either of maximums in (\ref{_9}) is known,  up to the moment, only for $s=1/2$. 

We stress that spin observables characterize the key intrinsic property of a quantum system  and the study of spin correlations in the context of Bell inequalities  violation under spin $s\geq 1$ measurements has been considered in the literature since the articles by Mermin et al. \cite{Mer:80, Mer:82} and is relevant at the present moment for many quantum applications, see  \cite{He.Dru.Rei:11} and references therein; moreover, last years for studies in particle physics 
\cite{bff,rev,sq}. 

In the present paper, for an arbitrary state of two spin $s\geq 1$ qudits, we analyze the maximum 
$\Upsilon_{chsh}^{(spin-s)}(\rho_{d\times d})$ of the absolute value of the CHSH expectation (\ref{0_0}) under local Alice and Bob spin-$s$ measurements.

For a state of two spin-$s$  qudits of dimension $d=2s+1\geq 2$, we introduce the notion of the spin-$s$ correlation matrix, which has dimension $3\times 3$ for all $s\geq \frac{1}{2}$; establish its relation to the general correlation $(d^{2}-1)\times(d^{2}-1)$ matrix of this state within the generalized Pauli representation and \emph{derive in terms of the spin-$s$ correlation matrix the explicit analytical expression
for the maximum} $\Upsilon_{chsh}^{(spin-s)}(\rho_{d\times d})$ under local Alice and Bob spin-$s$
measurements. 

As an application, we specify this general expression for Alice and Bob spin $s\geq \frac{1}{2}$ measurements in the two-qudit GHZ state, some nonseparable pure states and the two-qudit Werner state. 
We find that, under spin $s\geq 1$ measurements in each of these
nonseparable states, including the maximally entangled one, the CHSH
inequality is not violated. Moreover, unlike the case of spin $s=1/2$
measurements, where by (\ref{__8}), for every pure nonseparable two-qubit
state $|\psi_{2\times 2}\rangle$, the CHSH inequality is violated and parameter $M_{hor}(|\psi
_{2\times 2}\rangle )$ increases with a growth of the entanglement degree of this state, the situation under spin $s\geq 1$ measurements is
quite different -- namely, for a pure two-qudit state with a higher degree
of entanglement, its maximal CHSH expectation turns out to be less than
for a pure two-qudit state with lower entanglement and even for a separable one.

\section{The CHSH expectation under spin-$s$ measurements}
\label{section2}

Let $\rho _{d\times d}$ on $\mathbb{C}%
^{d}\otimes \mathbb{C}^{d}$  be an arbitrary state of two spin qudits, each of dimension  $d=2s+1\geq 2$, and 
\begin{align}
S_{r}& =r\cdot S,\text{ \ \ }r\cdot S:=\sum_{j=1,2,3}r_{j}S_{j},  \label{1} \\
r=&(r_{1},r_{2},r_{3}) \in \mathbb{R}^{3},\text{\ \ \ }\left\Vert
r\right\Vert _{\mathbb{R}^{3}}=1,  \notag
\end{align}%
be the quantum observable on $\mathbb{C}^{d}$ describing the
\textquotedblleft projection" onto a direction $r\in \mathbb{R}^{3}$ of the
qudit spin 
\begin{equation}
S=(S_{1},S_{2},S_{3}),\ \text{\ \ }S^{2}=S_{1}^{2}+S_{2}^{2}+S_{3}^{2}=s(s+1)%
\mathbb{I}_{\mathbb{C}^{d}},  \label{1.1}
\end{equation}
with the components 
\begin{align}
S_{1}& =\frac{1}{2}\sum_{m=1}^{d-1}\sqrt{m(d-m)}\left( |m\rangle \langle m+1|%
\text{ }+|m+1\rangle \langle m|\right) ,  \label{c} \\
S_{2}& =\frac{1}{2i}\sum_{m=1}^{d-1}\sqrt{m(d-m)}\left( |m\rangle \langle
m+1|\text{ }-|m+1\rangle \langle m|\right) ,  \notag \\
S_{3}& =\frac{1}{2}\sum_{m=1}^{d}\left( d+1-2m\right) |m\rangle \langle m|, \notag\\
S_{3}&|m\rangle=\bigl(s-(m-1)\bigr)|m\rangle, \ \ m=1,\dots,2s+1 ,
\notag
\end{align}%
that constitute particular cases of spin projections (\ref{1}) and  satisfy the relations \cite{Var.Mos.Khe:88} 
\begin{align}
\left[ S_{j},S_{k}\right] & =i\sum_{l}\varepsilon _{jkl}S_{l},\text{ \ \ }%
j,k,l=1,2,3,  \label{2} \\
\mathrm{tr}\{S_{j}S_{k}\}& =\frac{1}{3}s(s+1)(2s+1)\delta _{jk}=\frac{%
d(d^{2}-1)}{12}\delta _{jk},  \label{10_}
\end{align}%
where (i) $\varepsilon _{jkl}$ is the Levi-Civita symbol\footnote{The value of $\varepsilon _{jkl}$ is equal to unity if tuple $(j,k,l)$ constitutes an even permutation of $(1,2,3)$; minus unity if  $(j,k,l)$ is an odd permutation of $(1,2,3)$ and zero if values of some indices coincide.}; (ii) $\delta_{jk}$ is the Kronecker symbol\footnote{$\delta_{jk} =1$, for $j=k$, and $\delta_{jk}=0$, for $j\neq k$.}  and (iii) $\{\left\vert
m\right\rangle \in \mathbb{C}^{d},m=1,...,d\}$ is the computational
orthonormal basis in $\mathbb{C}^{d}.$  
By relations (\ref{1}) and (\ref{10_}) 
\begin{equation}
\mathrm{tr}\{S_{r}^{2}\}=\frac{s(s+1)(2s+1)}{3},\ \ r\in \mathbb{R}^{3},\ \
\left\Vert r\right\Vert _{\mathbb{R}^{3}}=1.  \label{2.2}
\end{equation}
Spin observables (\ref{1}) and  (\ref{c}) have the nondegenerate
eigenvalues $\left\lbrace-s,-(s-1),...,-1,\right.$ $\left.0,1,...,(s-1),s\right\rbrace$ including zero if spin $s$ is integer, and $\{-s,-(s-1),...,-\frac{1}{2},\frac{1}{2},...,(s-1),s\}$ if spin $s$ is half-integer. Hence, the eigenvalues of each spin-$s$ observable (\ref{1}) are in $[-s,s]$.

Consider a bipartite correlation scenario where each of two parties, say
Alice and Bob, performs measurements of two spin-$s$ observables (\ref{1})
in a two-qudit state $\rho _{d\times d}$. Let Alice measure spin observables 
$S_{a_{k}}=a_{k}\cdot S,$ $k=1,2$ and Bob -- spin observables $%
S_{b_{k}}=b_{k}\cdot S,$ $k=1,2.$ For this bipartite correlation scenario,
the CHSH operator $\mathcal{B}_{chsh}$ in (\ref{0_0}) 
takes the form 
\begin{equation}
\mathcal{B}_{chsh}(a_{1},a_{2};b_{1},b_{2})=S_{a_{1}}\otimes
(S_{b_{1}}+S_{b_{2}})+S_{a_{2}}\otimes (S_{b_{1}}-S_{b_{2}}),  \label{7}
\end{equation}%
so that its expectation
\begin{equation} \langle \mathcal{B}_{chsh}(a_{1},a_{2};b_{1},b_{2})%
\rangle _{\rho _{d\times d}}=\mathrm{tr}[\rho _{d\times d}\mathcal{B}%
_{chsh}(a_{1},a_{2};b_{1},b_{2})]\label{00}
\end{equation}
 in a state $\rho _{d\times d}$ is given by
\begin{equation}
\mathrm{tr}[\rho \{S_{a_{1}}\otimes (S_{b_{1}}+S_{b_{2}})\}]+\mathrm{tr}%
[\rho \{S_{a_{2}}\otimes (S_{b_{1}}-S_{b_{2}})\}].  \label{8}
\end{equation}%
By (\ref{1}) each
term in (\ref{8}) can be presented otherwise in the form
\begin{align}
\mathrm{tr}[\rho _{d\times d}\{S_{a_{k}}\otimes S_{b_{k}}\}]&
=\sum_{i,j}a_{k}^{(i)}\mathcal{Z}_{s}^{(ij)}(\rho _{d\times
d})b_{k}^{(j)}= \left(a_{k},\mathcal{Z}_{s}(\rho _{d\times d})b_{k}\right)_{\mathbb{R}^{3}} , 
\label{9}\\
a_{k},&b_{k} \in \mathbb{R}^{3},\text{\ \ \ }\left\Vert a_{k}\right\Vert _{%
\mathbb{R}^{3}},\left\Vert b_{k}\right\Vert _{\mathbb{R}^{3}}=1,  \notag
\end{align}%
where $\mathcal{Z}_{s}(\rho _{d\times d})$ is the $3\times 3$ matrix with
real elements 
\begin{align}
\mathcal{Z}_{s}^{(ij)}(\rho _{d\times d})&:=\mathrm{tr}[\rho _{d\times
d}\{S_{i}\otimes S_{j}\}]\in \mathbb{R}, \label{10}\text{ \ \ }i,j=1,2,3.  
\end{align}%
For $s=1$, this matrix was first introduced by Eq. (55) in \cite{Khr.Lou:18} for the analysis of the maximal violation of the original Bell inequality \cite{Bel:64} under spin-$1$ measurements. For an arbitrary spin $s\geq \frac{1}{2}$ we further refer to $\mathcal{Z}%
_{s}(\rho _{d\times d})$ as\textit{\ }\emph{the spin-}$\emph{s}$\emph{\
correlation matrix} of a state $\rho _{d\times d}.$

As it follows from the results in \cite{Khr.Lou:18} on violation of the original Bell inequality for spin $s=\frac{1}{2}$ and spin $s=1$  and  the below Theorem \ref{Theorem1} and Corollary \ref{Corollary1}, which are true  for all $s\geq\frac{1}{2}$, it is specifically the spin-$s$ correlation matrix of a two-qudit state that specifies  violation or nonviolation of a Bell inequality under spin-$s$ measurements on this state.

In order to evaluate the singular values of matrix (\ref{10}), we note that $\left\vert \mathrm{tr}[\rho _{d\times d}(A\otimes
B)]\right\vert \leq $ $\left\Vert A\right\Vert _{0}\left\Vert B\right\Vert
_{0}$ holds for any observables with operator norms $\left\Vert \cdot
\right\Vert _{0}$ and that, for the spin-$s$ observable (\ref{1}), the operator
norm $\left\Vert S_{r}\right\Vert _{0}=s$. This and relation (\ref{9}) imply
\begin{equation}
\left\vert \left( a_{k},\mathcal{Z}_{s}(\rho _{d\times d})b_{k}\right)_{\mathbb{R}^{3}}
\right\vert\leq s^{2},\;\;  a_{k},b_{k} \in \mathbb{R}^{3},\text{ \ }\left\Vert a_{k}\right\Vert _{%
\mathbb{R}^{3}},\left\Vert b_{k}\right\Vert _{\mathbb{R}^{3}}=1\ \label{12}.
\end{equation}%
By taking in (\ref{12}) the unit vectors 
\begin{equation}
 a_{k}=\frac{\mathcal{Z}_{s}(\rho
_{d\times d})n}{\left\Vert \mathcal{Z}_{s}(\rho _{d\times d})n\right\Vert _{%
\mathbb{R}^{3}}}\ ,\ \  \  \  b_{k}=n,   
\end{equation} where $n$ is an arbitrary unit vector in $%
\mathbb{R}^{3},$ we have $\left\Vert \mathcal{Z}_{s}(\rho _{d\times
d})n\right\Vert _{\mathbb{R}^{3}}\leq s^{2}$, so that
\begin{equation}
\left\Vert \mathcal{Z}_{s}(\rho _{d\times d})\right\Vert _{0}\leq s^{2}.\label{21}
\end{equation}%
The latter relation implies that the largest singular value of matrix $\mathcal{Z}_{s}(\rho
_{d\times d})$ is upper bounded by $s^{2}$.

From Eqs. (\ref{8})--(\ref{10}) it follows that in terms of the spin-$s$ correlation matrix (\ref{10}) the CHSH expectation (\ref{00}) takes the form
\begin{eqnarray}
&\left\langle \mathcal{B}_{chsh}(a_{1},a_{2};b_{1},b_{2})\right\rangle _{\rho
_{d\times d}} =\left(a_{1},\mathcal{Z}_{s}(\rho _{d\times
d})(b_{1}+b_{2})\right)_{\mathbb{R}^{3}} +\left(a_{2},\mathcal{Z}_{s}(\rho _{d\times
d})(b_{1}-b_{2})\right)_{\mathbb{R}^{3}}.\;\label{11}
\end{eqnarray}
Based on this relation, let us consider the maximum of the CHSH expectation
\begin{equation}
\Upsilon_{chsh}^{(spin-s)}(\rho_{d\times d}) : = \max_{\substack{ a_{k},b_{k}\in \mathbb{R}^{3},  \\ \left\Vert
a_{k}\right\Vert ,\left\Vert b_{k}\right\Vert =1}}\left\vert \left\langle 
\mathcal{B}_{chsh}(a_{1},a_{2};b_{1},b_{2})\right\rangle _{\rho _{d\times
d}}\right\vert   \label{13}
\end{equation}%
under local Alice and Bob spin-$s$ measurements in a state $\rho_{d\times d}$. 
The following general statement is proved in Appendix A.
\begin{theorem}
\label{Theorem1}
Let $\rho_{d\times d}$ be an arbitrary two-qudit state on $\mathbb{C}^{d}\otimes \mathbb{C}^{d},\ d=2s+1\geq 2$. Under Alice and Bob spin-$s$
measurements in this state, the maximum (\ref{13}) of the absolute value of the CHSH expectation (\ref{00}) is given by 
\begin{eqnarray}
\Upsilon_{chsh}^{(spin-s)}(\rho_{d\times d}) =2\sqrt{z_{s}^{2}(\rho _{d\times d})+\widetilde{z}%
_{s}^{2}(\rho _{d\times d})}, \text{\ \ \ \ } \label{22_new}
\end{eqnarray}%
where $z_{s}(\rho _{d\times d})$, $\widetilde{z}_{s}(\rho
_{d\times d})$ are two largest singular values of the spin-$s$ correlation matrix $\mathcal{Z}_{s}(\rho _{d\times d})$ of state $%
\rho _{d\times d}$, defined by (\ref{10}).
\end{theorem}

Since all spin-$s$ observables in relation (\ref{7}) have eigenvalues in $[-s,s]$, in the inequality  (\ref{0_33}) the maximal
absolute value of the CHSH expectation in an LHV case should be replaced by $%
2s^{2}$. This, Theorem \ref{Theorem1}  and Eq.(\ref{21}) imply.

\begin{corollary}
\label{Corollary1}
For an arbitrary state $\rho _{d\times d}$, the ratio of the quantum maximum (\ref{22_new}) to the
maximal CHSH value in an LHV case is given by
\begin{align}
\gamma _{s}(\rho _{d\times d})& =\frac{1}{2s^{2}}\Upsilon_{chsh}^{(spin-s)}(\rho_{d\times d}) =\frac{1}{s^{2}}\sqrt{z_{s}^{2}(\rho _{d\times d})+\widetilde{z%
}_{s}^{2}(\rho _{d\times d})}  \label{23}
\end{align}%
and, in view of (\ref{21}), is upper bounded by the Tsirelson \cite{7,8} bound $%
\sqrt{2}$. A two-qudit state $\rho _{d\times d}$ violates the
CHSH inequality under Alice and Bob spin-$s$ measurements iff its
parameter $\gamma _{s}(\rho _{d\times d})>1.$ 
\end{corollary}

In what follows, we refer to parameter (\ref{23}) as \emph{the spin-$s$ CHSH parameter of a two-qudit state}.

For spin $s=1/2$, the spin-$\frac{1}{2}$ observables  (\ref{1}) are equal to 
$\frac{1}{2}\sigma _{r}$; the spin-$1/2$ correlation matrix coincides $
\mathcal{Z}_{s=1/2}(\rho _{2\times 2})=\frac{1}{4} T_{d=2}(\rho _{2\times 2})$, up to  the coefficient $\frac{1}{4}$, with the general correlation matrix (\ref{0_5}) of a two-qubit state and the spin-$s$ CHSH parameter $\gamma _{s=1/2}(\rho _{2\times 2})$ equals
exactly to the two-qubit Horodecki parameter (\ref{---7}).

For an arbitrary two-qudit state $\rho _{d\times d}$, $d=2s+1\geq 3$,  the
relation between the spin $s\geq 1$ correlation $3\times 3$ matrix $%
\mathcal{Z}_{s}(\rho _{d\times d})$ and the (general) correlation $\left(
d^{2}-1\right) \times \left( d^{2}-1\right) $ matrix $T_{d}(\rho _{d\times
d})$ is found in Section \ref{section4}.

\section{Spin-$s$ measurements in the two-qudit GHZ state}
\label{section3}
As an application of our general result (\ref{22_new}) in Theorem \ref{Theorem1}, let us calculate the spin-$s$ correlation
matrix $\mathcal{Z}_{s}(|\psi _{d\times d}^{(ghz)}\rangle )$ of the
GHZ state 
\begin{equation}
|\psi _{d\times d}^{(ghz)}\rangle =\frac{1}{\sqrt{d}}\sum_{m=1}^{d}|m\rangle
\otimes |m\rangle ,  \label{27}
\end{equation}%
which is pure and maximally entangled. From (\ref{10}) it follows that all
elements of matrix $\mathcal{Z}_{s}(|\psi _{d\times d}^{(ghz)}\rangle )$ have the
form 
\begin{equation}
\mathcal{Z}_{s}^{(ij)}(|\psi _{d\times d}^{(ghz)}\rangle )=\frac{1}{d}%
\sum_{m,m^{\prime }}\langle m^{\prime }|S_{i}|m\rangle \langle m^{\prime
}|S_{j}|m\rangle .  \label{27.1}
\end{equation}%
Taking into account in Eq. (\ref{27.1}) that by (\ref{c}) the elements of matrices $%
S_{1}$, $S_{3}$ are real and symmetric, while $\langle m^{\prime
}|S_{2}|m\rangle =-\langle m|S_{2}|m^{\prime }\rangle $, we derive 
\begin{align}
\mathcal{Z}_{s}^{(1j)}(|\psi _{d\times d}^{(ghz)}\rangle )& =\frac{1}{d}%
\sum_{m,m^{\prime }}\langle m^{\prime }|S_{1}|m\rangle \langle m^{\prime
}|S_{j}|m\rangle  \label{ghz} \\
& =\frac{1}{d}\sum_{m}\langle m|S_{1}S_{j}|m\rangle =\frac{1}{d}\mathrm{tr}%
[S_{1}S_{j}]=\frac{d^{2}-1}{12}\delta _{1j},  \notag \\
\mathcal{Z}_{s}^{(2j)}(|\psi _{d\times d}^{(ghz)}\rangle )& =\frac{1}{d}\sum
\langle m^{\prime }|S_{2}|m\rangle \langle m^{\prime }|S_{j}|m\rangle  \notag
\\
& =-\frac{1}{d}\sum_{m}\langle m|S_{2}S_{j}|m\rangle =-\frac{1}{d}\mathrm{tr}%
[S_{2}S_{j}]=-\frac{d^{2}-1}{12}\delta _{2j},  \notag \\
\mathcal{Z}_{s}^{(3j)}(|\psi _{d\times d}^{(ghz)}\rangle )& =\frac{1}{d}\sum
\langle m^{\prime }|S_{3}|m\rangle \langle m^{\prime }|S_{j}|m\rangle  \notag
\\
& =\frac{1}{d}\sum_{m}\langle m|S_{3}S_{j}|m\rangle =\frac{1}{d}\mathrm{tr}%
[S_{3}S_{j}]=\frac{d^{2}-1}{12}\delta _{3j}.  \notag
\end{align}%
Therefore, for the two-qudit GHZ state (\ref{27}), the spin-$s$ correlation
matrix is diagonal and given by
\begin{equation}
\mathcal{Z}_{s}(|\psi _{d\times d}^{(ghz)}\rangle )=\frac{d^{2}-1}{12}%
\begin{pmatrix}
1 & 0 & 0 \\ 
0 & -1 & 0 \\ 
0 & 0 & 1%
\end{pmatrix}%
,  \label{x}
\end{equation}%
and the spin-$s$ CHSH parameter (\ref{23}) equals to
\begin{equation}
\gamma _{s}(|\psi _{d\times d}^{(ghz)}\rangle )
=\frac{\sqrt{2}}{3}\frac{d+1}{d-1}=\frac{\sqrt{2}}{3}\frac{s+1}{s},\text{ \ \ }\text{}\  s\geq \frac{1}{2}.  \label{xx}
\end{equation}
For spin $s=\frac{1}{2}$ (dimension $d=2),$ the two-qubit GHZ state $|\psi
_{2\times 2}^{(ghz)}\rangle $ constitutes the Bell state $|\beta
_{00}\rangle $ and relation (\ref{xx}) reads $\gamma _{s=1/2}(|\beta
_{00}\rangle )=\sqrt{2}$. 

However, for all spins $s\geq 1$ $(d\geq 3),$ from Eq. (\ref{xx}) it
follows that parameter $\gamma _{s\geq 1}(|\psi _{d\times d}^{(ghz)}\rangle
)<1.$ By Corollary \ref{Corollary1} the latter relation  means that, under Alice and Bob measurements of spin 
$s\geq 1$ in the two-qudit GHZ state the CHSH
inequality is not violated.

We note that, for spin $s=1$, parameter $\gamma _{s=1}(|\psi _{3\times
3}^{(ghz)}\rangle )=\frac{2}{3}\sqrt{2}$ and this value coincides exactly
with the one presented by Eq. (63) in \cite{Lou:20} for the maximal
violation by the two-qutrit GHZ state $|\psi _{3\times 3}^{(ghz)}\rangle $
of the CHSH inequality under local Alice and Bob measurements of traceless qutrit
observables not necessarily spins. 

The latter means that, under
local Alice and Bob measurements of traceless observables in the
two-qutrit GHZ state $|\psi _{3\times 3}^{(ghz)}\rangle ,$ the maximal value
of the CHSH expectation (\ref{0_0}) is attained at spin-$1$ observables.

\section{Relation between the spin-$s$ correlation matrix of a two-qudit state and its general
correlation matrix}
\label{section4}

As mentioned above in Section \ref{section2}, for spin $s=\frac{1}{2},$ the spin
operators (\ref{c}) are proportional to the Pauli operators $%
\sigma_{i}$ as $S_{i}=\frac{1}{2}\sigma_{i}$, so that, for a two-qubit state $%
\rho_{2\times2},$ the spin-$\frac{1}{2}$ correlation matrix coincides $
\mathcal{Z}_{s=1/2}(\rho_{2\times 2})=\frac{1}{4}T_{d=2}(\rho_{2\times2})$, up to
the coefficient, with the correlation matrix (\ref{0_5}), introduced in \cite{Hor.Hor.Hor:95}.

For an arbitrary state $\rho _{d\times d},$ of two spin $s\geq1$ qudits of dimension $d=2s+1>2$, let us now find a relation between its spin-$s$ correlation matrix $
Z_{s}(\rho _{d\times d})$ and its general correlation matrix $T_{d}(\rho
_{d\times d}),$ defined in the frame of the generalized Pauli representation in 
\cite{Lou:20, Lou.Kul:21} via the relation%
\begin{align}
T_{d}^{(ij)}(\rho _{d\times d})&:=\mathrm{tr}\left[ \rho _{d\times d}\left(
\Lambda _{d}^{(i)}\otimes \Lambda _{d}^{(j)}\right) \right]\ , \ 
i,j=1,\dots ,d^{2}-1.  \label{23.3} 
\end{align}
Here, $\Lambda _{d}:=(\Lambda _{d}^{(1)},...,\Lambda _{d}^{(d^{2}-1)})$ is a
tuple of the generalized Gell-Mann operators, which are Hermitian,
traceless, satisfy the relation \textrm{tr}$[\Lambda _{d}^{(i)}\Lambda
_{d}^{(j)}]=2\delta _{ij}$ and constitute the higher-dimensional extensions
of the Pauli operators on $\mathbb{C}^2$ and the Gell-Mann operators  on $\mathbb{C}^3$, see \cite{Lou:20, Lou.Kul:21} for details.

Let $S_{k}=\sum_{j}n_{k}^{(j)}\Lambda _{d}^{(j)}$ be the decomposition
of the spin components in (\ref{c}) via the operator basis $\{%
\mathbb{I}_{d},\Lambda _{d}\}$ in the frame of the general Bloch formalism\ 
\cite{Lou.Kul:21}. Here,%
\begin{equation}
n_{k}=\frac{1}{2}\mathrm{tr}[S_{k}\Lambda _{d}]\in \mathbb{R}^{d^{2}-1},%
\text{ \ \ }k=1,2,3,  \label{nn}
\end{equation}%
are the Bloch vectors of spin components $S_{k},$ $k=1,2,3,$ and these Bloch
vectors satisfy the relation 
\begin{equation}
\mathrm{tr}[S_{k}S_{j}]=2( n_{k},n_{j})_{\mathbb{R}^{d^{2}-1}} ,\text{ \ \ }k,j=1,2,3.
\label{23.5}
\end{equation}%
Therefore, by (\ref{10_}) and (\ref{23.5})
\begin{equation}
( n_{k},n_{j})_{\mathbb{R}^{d^{2}-1}}  =\frac{d(d^{2}-1)}{24}\delta _{kj},\text{ \ \ }%
k,j=1,2,3.\label{new2}
\end{equation}

Substituting decomposition $S_{k}=\sum_{j}n_{k}^{(j)}\Lambda _{d}^{(j)}$ into relation (\ref{10}) and taking into account Eqs. (\ref{nn}) and (\ref{23.5}), we find for an arbitrary two-qudit state 
$\rho _{d\times d}$ the following relation between its spin-$s$ correlation
matrix $\mathcal{Z}_{s}\mathcal{(}\rho _{d\times d})$ and its general
correlation matrix $T_{d}(\rho _{d\times d}).$

\begin{theorem}
\label{Theorem2}
For an arbitrary two-qudit state $\rho _{d\times d},$  the elements of its
spin-$s$ correlation matrix $\mathcal{Z}_{s}\mathcal{(}\rho _{d\times d})$
defined by relation (\ref{10}) and its general correlation matrix $T_{d}(\rho
_{d\times d})$ in (\ref{23.3}) satisfy the relation:%
\begin{equation}
\mathcal{Z}_{s}^{(ij)}\mathcal{(}\rho _{d\times d})=\left( n_{i},T_{d}%
\mathcal{(}\rho _{d\times d})n_{j}\right)_{\mathbb{R}^{d^{2}-1}}.\label{23.4}
\end{equation}
\end{theorem}

By applying (\ref{23.4}), let us, for example, find the spin-$s$ correlation
matrix $\mathcal{Z}_{s}(|\psi _{d\times d}^{(ghz)}\rangle )$ for
the two-qudit GHZ state (\ref{27}). 

For the GHZ state,  the general correlation matrix $T_{d}(|\psi _{d\times
d}^{(ghz)}\rangle) $,  found in \cite{Lou:20} by Eq. (53),  constitutes the
block diagonal matrix, composed of three diagonal matrices $T_{d}^{(sym)},$ $%
T_{d}^{(asym)}$and $T_{d}^{(diag)}$ where: (i) matrix $T_{d}^{(sym)}$ has $%
d(d-1)/2$ nonzero elements, each equal to $2/d$ and corresponding due to (%
\ref{23.3}) to the symmetric Gell-Mann operators $\Lambda _{d}^{(sym)};$
(ii) matrix $T_{d}^{(asym)}$ has also $d(d-1)/2$ nonzero diagonal elements,
each equal to $(-2/d)$, and corresponding due to (\ref{23.3}) to the
antisymmetric Gell-Mann operators $\Lambda _{d}^{(asym)};$ and (iii) matrix $%
T_{d}^{(diag)}$ has $(d-1)$ elements, corresponding due to (\ref{23.3}) to
the diagonal Gell-Mann operators $\Lambda _{d}^{(diag)}$ and each equal to $%
2/d$.

In view of relations (\ref{c}) for spins $S_{j},$ $j=1,2,3,$ and
the explicit form of the generalized Gell-Mann operators presented by Eqs. (4)--(6) in \cite{Lou:20}, the Bloch vector $n_{1}=\frac{1}{2}\mathrm{tr}%
[S_{1}\Lambda _{d}]$ has only nonzero $d(d-1)/2$ components, corresponding
to the symmetric generalized Gell-Mann components of the tuple $\Lambda _{d}$%
, the Bloch vector vector $n_{2}=\frac{1}{2}\mathrm{tr}[S_{2}\Lambda _{d}]$
-- only nonzero $d(d-1)/2$ components, corresponding to the antisymmetric
generalized Gell-Mann components of the tuple $\Lambda _{d}$, and the
Bloch vector $n_{3}=\frac{1}{2}\mathrm{tr}[S_{3}\Lambda _{d}]$ - only $(d-1)$
nonzero components corresponding to the diagonal components of $\Lambda _{d}.$ 

All this and Eq. (\ref{new2}) imply that, for the spin-$s$ correlation matrix $\mathcal{Z}_{s}%
\mathcal{(}|\psi _{d\times d}^{(ghz)}\rangle )$ of the two-qudit GHZ state, the expression, calculated via relation
(\ref{23.4}), coincides with the above expression (\ref{x}), determined via definition (\ref{10}) of this correlation matrix. 

\section{Spin correlation matrix for an arbitrary two-qudit state}

In this section, we specify the elements of the spin-$s$ correlation matrix (\ref{10}) for an arbitrary two-qudit state $\rho_{d\times d}$, pure and mixed. 
Note that if a two-qudit  state is invariant under the permutation of the Hilbert spaces in the tensor product $\mathbb{C}^{d}\otimes\mathbb{C}^{d}$, then by relation $(\ref{10}$) its spin-$s$
correlation matrix is symmetric.

The spectral decomposition $\rho_{d\times d} =\sum_{n}\lambda_{n}|\Psi_{n}\rangle
\langle\Psi_{n}|,$ $\lambda_{n}>0,$ $\sum\lambda_{n}=1,$ of a two-qudit state and the representation of each of its eigenvectors $|\Psi_{n}%
\rangle=\sum\eta_{mk}^{(n)}|m\rangle\otimes|k\rangle$ in the computational basis in
$\mathbb{C}^{d}\otimes\mathbb{C}^{d}$ imply 
\begin{align}
\rho_{d\times d} &  =\sum\zeta_{mm^{\prime},kk^{\prime}}|mk\rangle\langle
m^{\prime}k^{\prime}|,\label{22} \\
&
\zeta_{mm^{\prime},kk^{\prime}}=\langle
mk|\rho_{d\times d}|m^{\prime}k^{\prime}\rangle=\sum_{n}\lambda_{n}\eta
_{mk}^{(n)}\left(  \eta_{m^{\prime}k^{\prime}}^{(n)}\right)  ^{\ast},\nonumber\\
&
\zeta_{mm^{\prime},kk^{\prime}}^{\ast}  =\zeta_{m^{\prime}m,k^{\prime}%
k},\text{ \ \ }\sum_{m,k}\zeta_{mm,kk}=1.\nonumber
\end{align}
This and relations (\ref{c}), (\ref{10}) imply the following expressions for the elements  of the spin-$s$ correlation matrix for state $\rho_{d\times d}$.  
In the first line:
\begin{align}
\mathcal{Z}^{(11)}_{s}(\rho_{d\times d}) &  =\frac{1}{2}\sum_{m,k=1}^{d-1}\text{
}\sqrt{mk(d-m)(d-k)}\mathrm{Re}\left[  \text{ }\zeta_{m(m+1),k(k+1)}%
+\zeta_{m(m+1),(k+1)k}\right]  ,\label{Z11d}\\
\mathcal{Z}^{(12)}_{s}(\rho_{d\times d}) &  =\frac{1}{2}\sum_{m,k=1}^{d-1}\text{
}\sqrt{mk(d-m)(d-k)}\mathrm{Im}\left[  \text{ }\zeta_{m(m+1),k(k+1)}%
+\zeta_{(m+1)m,k(k+1)}\right]  ,\nonumber\\
\mathcal{Z}^{(13)}_{s}(\rho_{d\times d}) &  =\frac{1}{2}\sum_{m,k=1}^{d}\text{
}\sqrt{m(d-m)}\left(  d+1-2k\right)  \mathrm{Re}\left[  \zeta_{(m+1)m,kk}%
\right]  .\nonumber
\end{align}

In the second line:
\begin{align}
\mathcal{Z}^{(21)}_{s}(\rho_{d\times d}) &  =\frac{1}{2}\sum_{m,k=1}^{d-1}\text{
}\sqrt{mk(d-m)(d-k)}\text{ }\mathrm{Im}\left[  \zeta_{m(m+1),k(k+1)}
+\zeta_{m(m+1),(k+1)k}\right]  ,\label{Z21d}\\
\mathcal{Z}^{(22)}_{s}(\rho_{d\times d}) &  =\frac{1}{2}\sum_{m,k=1}^{d-1}\text{
}\sqrt{mk(d-m)(d-k)}\text{ }\mathrm{Re}\left[  \zeta_{(m+1)m,k(k+1)}%
-\zeta_{(m+1)m,(k+1)k}\right]  ,\nonumber\\
\mathcal{Z}^{(23)}_{s}(\rho_{d\times d}) &  =\frac{1}{2}\sum_{m,k=1}^{d}\text{
}\sqrt{m(d-m)}\left(  d+1-2k\right)  \mathrm{Im}\left[  \zeta_{m(m+1),kk}%
\right]  ,\nonumber
\end{align}
and in the third line:
\begin{align}
\mathcal{Z}^{(31)}_{s}(\rho_{d\times d}) &  =\frac{1}{2}\sum_{m,k=1}^{d}\text{
}\left(  d+1-2m\right)\sqrt{k(d-k)} \mathrm{Re}\left[  \zeta_{mm,(k+1)k}
\right]  ,\label{Z31d}\\
\mathcal{Z}^{(32)}_{s}(\rho_{d\times d}) &  =\frac{1}{2}\sum_{m,k=1}^{d}\text{
}\left(  d+1-2m\right)\sqrt{k(d-k)}  \mathrm{Im}\left[  \zeta_{mm,k(k+1)}
\right]  ,\nonumber\\
\mathcal{Z}^{(33)}_{s}(\rho_{d\times d}) &  =\frac{1}{4}\sum_{m,k=1}^{d}%
(d+1-2m)(d+1-2k)\zeta_{mm,kk}.\ \nonumber
\end{align}

From relations (\ref{Z11d})--(\ref{Z31d}) it, in particular, follows that if, for a two-qudit  state, coefficients $\zeta_{mm^{\prime},kk^{\prime}}$ in  decomposition (\ref{22}) are real, then the spin-$s$ correlation matrix of this state takes the form 
\begin{equation}
\mathcal{Z}_{s}(\rho_{d\times d})=\left(
\begin{array}
[c]{ccc}%
\mathcal{Z}_{11} & 0 & \mathcal{Z}_{13}\\
0 & \mathcal{Z}_{22} & 0\\
\mathcal{Z}_{31} & 0 & \mathcal{Z}_{33}%
\end{array}
\right),  
\end{equation}
and the singular values of this matrix needed for finding the spin-$s$ CHSH parameter of this state via relation (\ref{23}) can be easily calculated. 

\section{Examples}
Besides the spin-$s$ CHSH parameter (\ref{xx}) for the GHZ state, which we calculated in Section \ref{section3} based only on definition (\ref{23}) of this parameter, let us further specify by expressions (\ref{Z11d})--(\ref{Z31d}) the spin-$s$ CHSH parameter for some other entangled two-qudit states, pure and mixed.

\subsection{For pure two-qudit states with  diagonal spin-$s$ correlation matrices}
Consider a pure two-qudit state with vector $|\psi_{d\times
d}\rangle \in \mathbb{C}^{d}\otimes \mathbb{C}^{d}$ admitting the Schmidt decomposition of the
form%
\begin{equation}
|\psi_{d\times d}\rangle =\sum_{m=1}^{d}\sqrt{\mu _{m}}%
|m\rangle \otimes |m\rangle ,\label{y}
\end{equation}%
where $\mu _{m}\geq 0,$ $\sum_{i}\mu _{m}=1.$   The two-qudit GHZ state (\ref{27}) is a
particular case of (\ref{y}) if $\mu _{m}=\frac{1}{\sqrt{d}},  m=1,...,d.$

For the pure state (\ref{y}), the coefficients in decomposition (\ref{22})  are given by $\zeta_{m m^{\prime}, kk^{\prime}}=\sqrt{\mu_{m}\mu_{m}^{\prime
}}\delta_{mk}\delta_{m^{\prime}k^{\prime}}$, where $\delta_{mk}$ is the Kronecker symbol. This and relations (\ref{Z11d})--(\ref{Z31d}) imply that all the off-diagonal elements of matrix $\mathcal{Z}_{s}
(|\psi_{d\times d}\rangle )$ are equal to zero while the diagonal
elements have the form (see in Appendix B) 
\begin{align}
\mathcal{Z}_{s}^{(11)}(|\psi_{d\times d}\rangle ) &=-\mathcal{Z} 
_{s}^{(22)}(|\psi _{d\times d}\rangle )   =\sum_{k=1}^{2s}k\left(s-\frac{k-1}{2}\right)\sqrt{\mu _{k}\mu _{k+1}},  \label{y2} \\
\mathcal{Z}_{s}^{(33)}(|\psi_{d\times d}\rangle )&
=\sum_{k=1}^{2s+1}\left( s-(k-1)\right) ^{2}\mu _{k},  \notag
\end{align}
so that, for state (\ref{y}), the spin-$s$ correlation matrix $\mathcal{Z}_{s}(|\psi_{d\times d}\rangle )$ has the singular values $\ $ $\sum_{k=1}^{2s}k\left(s-\frac{k-1}{2}\right)\sqrt{\mu _{k}\mu _{k+1}}\ $ of multiplicity $2$
and $ \sum_{k=1}^{2s+1}\left( s-(k-1)\right) ^{2}\mu _{k} $
of multiplicity $1$.

Substituting these singular values into relation (\ref{23}) gives the
explicit value of the spin-$s$ CHSH parameter\ for each specific state (\ref{y}).

As an example, consider for $d>2$ ($s\geq 1$)  the case, where in (\ref{y}), there are only  two nonzero terms: $\mu _{k}=\mu _{n}=%
\frac{1}{2},$ $n> k+1.$ For this particular case, state   (\ref{y}) reduces to
\begin{align}
|\psi_{d\times d}(k,n)\rangle =\frac{1}{\sqrt{2}}\left(
|k\rangle \otimes |k\rangle +|n\rangle \otimes |n\rangle \right)   \label{pp}
\end{align}%
and by Eqs. (\ref{23}) and (\ref{y2}) its spin-$s$ CHSH parameter is given by

\begin{align}
\gamma _{s}(|\psi_{d\times d}(k,n)\rangle ) &=\frac{1}{2s^{2}}\left\{ \left( s-(k-1)\right) ^{2}+\left( 
s-(n-1)\right) ^{2}\right\} \  \  \  \ \  \label{w}
\end{align}
and, for all $s\geq 1$, \begin{equation}
\gamma _{s}(|\psi_{d\times d}(k,n)\rangle )\leq1,\label{k}
\end{equation}
where the equality in (\ref{k}) holds if $k=1$ and $n=d$. 
By Corollary \ref{Corollary1} relation (\ref{k}) means that, under spin $s\geq 1$ measurements in the nonseparable two-qudit state (\ref{pp}), the CHSH inequality is not violated.

If we take a pure separable state, say $|n\rangle \otimes |n\rangle $, then
by (\ref{y2}) and (\ref{23}) its spin-$s$ CHSH parameter 
\begin{equation}
\gamma _{s}(|nn\rangle )=\left( 1-\frac{n-1}{s}\right) ^{2}\leq 1,\text{ \ \ 
}s\geq \frac{1}{2},  \label{ww}
\end{equation}%
and attains the value $1$ for $n=1$ or $n=d.$ 

The latter implies that,  under high ($s\geq 1$) spin-$s$ measurements in
pure separable states $|11\rangle $ and $|dd\rangle $, the spin-$s$ CHSH parameters of these states are equal to that of  the nonseparable state  $|
\psi_{d\times d}(1,d)\rangle,$ moreover, are greater than the spin-$s$ CHSH
parameter (\ref{xx}) of two-qudit GHZ state, which is maximally entangled.

\begin{remark}
\label{Remark1}
For the nonseparable two-qudit
state (\ref{pp}), the concurrence $\mathrm{C}(|%
\psi_{d\times d}(k,n)\rangle )=1$ for  all $d>2$ and this value is less than the value of the concurrence $\mathrm{C}(|\psi _{d\times d}^{(ghz)}\rangle )=\sqrt{%
\frac{2(d-1)}{d}}$ of the maximally entangled two-qudit GHZ state (\ref{27}).  However, unlike the monotonically increasing dependence (\ref{__8}) on the concurrence of the maximal CHSH expectation (\ref{0_4}) under spin-$1/2$ measurements in a pure two-qubit state, in case of spin $s\geq1$ measurements, by Eqs. (\ref{xx}), (\ref{k}) and (\ref{ww}) the maximal CHSH expectation 
 in the two-qudit GHZ state is less than the maximal CHSH expectation in the nonseparable state $|\psi _{d\times
d}(1,d)\rangle$ and even in separable states $|11\rangle $ and $%
|dd\rangle $.
\end{remark}

\subsection{For the two-qudit Werner state}

Let us calculate the spin-$s$ CHSH parameter (\ref{23}) of the two-qudit Werner state \cite{Wer:89}
\begin{align}
\rho _{d,\Phi }^{(wer)}=\frac{d-\Phi }{d(d^{2}-1)}&\mathbb{I}_{\mathbb{C}%
^{d}\otimes \mathbb{C}^{d}}+\frac{d\Phi -1}{d(d^{2}-1)}V_{d}\label{g} ,\ \ \text{}\
\Phi \ \in \left[ -1,1\right] ,
\end{align}
where $V_{d}(\psi _{1}\otimes \psi _{2}):=\psi _{2}\otimes \psi _{1}$ is the
permutation operator on $\mathbb{C}^{d}\otimes \mathbb{C}^{d}$. Recall \cite{Wer:89} that the Werner
state $\rho _{d,\Phi }^{(wer)}$ is separable iff $\Phi \ \in \ \left[ 0,1%
\right] $ and nonseparable otherwise, also, under projective measurements of Alice
and Bob, the nonseparable Werner state $\rho _{d,\Phi }^{(wer)}$ admits an LHV model for all $\Phi \in \lbrack -1+\frac{d+1}{d^{2}},0).$

For the Werner state $\rho_{d,\Phi}^{(wer)},$ the coefficients in decomposition (\ref{22}) are given by
\begin{equation}
\zeta_{mm^{\prime},kk^{\prime}}=\frac{d-\Phi}{d(d^{2}-1)}\delta_{m,m^{\prime}%
}\delta_{k,k^{\prime}}+\frac{d\Phi-1}{d(d^{2}-1)}\delta_{mk^{\prime}}%
\delta_{m^{\prime}k}.\label{coeff_werner}%
\end{equation}
From relations (\ref{coeff_werner}) and (\ref{Z11d})--(\ref{Z31d})
it follows that the spin-$s$
correlation matrix of the Werner state $\rho _{d,\Phi }^{(wer)}$ has the form (see Appendix in C)
\begin{equation}
\mathcal{Z}_{s}(\rho _{d,\Phi }^{(wer)})=\frac{d\Phi -1}{12}\mathbb{I}_{%
\mathbb{R}^{3}},\text{ \ \ }d=2s+1\geq 2,
\end{equation}%
and, therefore, the
singular value $|d\Phi -1|/12$ of multiplicity $3$. 

This and relation (\ref{23}) imply that, for the Werner state $
\rho _{d,\Phi }^{(wer)}$,  the spin-$s$ CHSH parameter 
\begin{eqnarray}
\gamma _{s}(\rho _{d,\Phi }^{(wer)}) &=&\frac{\sqrt{2}}{3}\frac{|d\Phi -1|}{(d-1)^{2}}, \; \;\;\;\; d\geq 2. \label {yy}
\end{eqnarray}

For spin $s=\frac{1}{2}$ $(d=2),$ parameter $\gamma _{s=1/2}(\rho _{2,\Phi
}^{(wer)})=\frac{\sqrt{2}}{3}|2\Phi -1|$ and this 
coincides with the value of the Horodecki parameter $
M_{hor}(\rho _{2,\Phi }^{(wer)})$ for the two-qubit Werner state, which was found in \cite{Hor.Hor.Hor:95}. 

For all $\Phi \in \lbrack -1,-\frac{3}{4}\sqrt{2}+\frac{1}{%
2})$, the spin-$1/2$ parameter $\gamma _{s=1/2}(\rho _{2,\Phi }^{(wer)})>1$,
so that by Corollary \ref{Corollary1}, under spin-$1/2$ measurements in the nonseparable two-qubit Werner states $%
\rho _{2,\Phi }^{(wer)}$ with these $\Phi ,$ the CHSH inequality is violated.
If $\Phi \in \lbrack -\frac{3}{4}\sqrt{2}+\frac{1}{2},$ $0),$ then
parameter $\gamma _{s=1/2}(\rho _{2,\Phi }^{(wer)})\leq 1$, therefore, by Corollary \ref{Corollary1} the corresponding nonseparable two-qubit Werner states $\rho _{2,\Phi }^{(wer)}$ do not violate the CHSH inequality under spin-$\frac{1}{2}$ measurements and this is
consistent with the fact that, as found in \cite{Wer:89}, under all
bipartite projective measurements, the nonseparable two-qubit Werner state
with $\Phi \in \lbrack -\frac{1}{4},0)$ admits an LHV model.    

For spin $s\geq 1$ $(d\geq 3)$, the spin-$s$ CHSH
parameter $\gamma _{s\geq 1}(\rho _{d,\Phi }^{(wer)})<1$
for all $\Phi \in \lbrack -1,1]$. Therefore, by Corollary \ref{Corollary1}, under Alice and Bob spin-$s$ 
measurements in any nonseparable two-qudit Werner state $\rho _{d,\Phi}^{(wer)}, d\geq 3$,  the CHSH inequality is not violated. The latter agrees with Theorems 2 and 3 in \cite{Lou:05}, where it is proved that, for all $d\geq 3$,  every nonseparable two-qudit Werner
state does not violate the CHSH inequality under Alice and Bob
measurements of traceless qudit observables, not necessarily qudit spins.

\section{Conclusion}

In the present article, 
for an arbitrary state $\rho _{d\times d}$  of two spin-$s$ qudits of dimension $d=2s+1\geq 2$, we introduce the notion of the  spin-$s$ correlation
matrix (\ref{10}), having dimension $3\times3$  for all $d\geq2$, and derive (Theorem \ref{Theorem1}) in
terms of this 
matrix the explicit analytical expression for the maximum of the absolute value of the CHSH expectation (\ref{0_0}) under local Alice and Bob spin-$s$ measurements in this state.  In view of relations (\ref{0_4}) and (\ref{_9}), this explicit analytical expression specifies for all $s\geq\frac{1}{2}$ the tight lower bound on the maximum of the CHSH expectation (\ref{0_0}) over all local Alice and Bob qudit observables. 

In Theorem \ref{Theorem2}, we establish the exact relation between the  spin-$s$ correlation matrix (\ref{10}) of an arbitrary two-qudit state and the general correlation matrix (\ref{23.3}) of this state, defined within its generalized Pauli representation \cite{Lou.Kul:21}.

Calculating the spin-$s$ correlation matrix for the two-qudit GHZ state (\ref{27}), some pure nonseparable two-qudit states (\ref{y}) and the
two-qudit Werner state (\ref{g}), we find that, under high-spin ($
s\geq 1$) measurements in each of these nonseparable states,
including the maximally entangled one, \emph{the CHSH inequality is not
violated.}

Moreover, unlike the case of spin-$1/2$ measurements where each pure
nonseparable two-qubit state violates the CHSH inequality and its maximal CHSH expectation increases monotonically with a growth of its entanglement, the
situation under spin $s\geq 1$) measurements in a two-qudit state is quite different. As we
discuss in Remark \ref{Remark1}, for a pure two-qudit state with a higher degree of  
entanglement, the maximal CHSH expectation under spin $s\geq1$ measurements turns out to be less than that for a pure two-qudit state with a lower entanglement and even for some separable states.


The new results of the present article provide a tool to analyze a degree of violation of the CHSH inequality under local Alice and Bob spin $s\geq 1$ measurements in a nonseparable two-qudit state. The solution of this problem, which is relevant for a variety of quantum technologies tasks, also, in particle physics, contributes to a deeper understanding of Bell nonlocality \cite{Bru.etal:14, Lou:17} exhibited by a nonseparable two-qudit state via the CHSH inequality under spin $s\geq 1$ measurements. Based on the results of the present article, we have, in particular, investigated this issue in our subsequent research \cite{Han.Lou:25} on the spin-$1$ case.

\section{Acknowledgments}

The study by E.R. Loubenets in Sections 1, 2, 3 and 4 of this work was supported
by the Russian Science Foundation under the Grant № 24-11-00145,  https://rscf.ru/en/project/24-11-00145/ and performed at the Steklov Mathematical Institute of Russian Academy of Sciences. The study by E.R. Loubenets and L. Hanotel in Sections 5, 6 and 7 was performed at the HSE University. The authors are grateful to Professor A.V.  Borisov for useful discussions.
\vspace{0.4cm}

\section*{Appendix A}
\begin{proof}[Proof of Theorem \ref{Theorem1}] From representation (\ref{11}) of the CHSH expectation via the sum of two scalar products it follows that in (\ref{13}) the maximum over all unit vectors $a_{k}\in \mathbb{R%
}^{3},$ $k=1,2,$ is attained at the unit vectors 
\begin{eqnarray}
\widetilde{a}_{1} &=&\frac{\mathcal{Z}_{s}(\rho _{d\times d})\left(
b_{1}+b_{2}\right) }{\left\Vert \mathcal{Z}_{s}(\rho _{d\times d})\left(
b_{1}+b_{2}\right) \right\Vert _{\mathbb{R}^{3}}}\in \mathbb{R}^{3},
\label{14} \\
\text{\ }\widetilde{a}_{2} &=&\frac{\mathcal{Z}_{s}(\rho _{d\times d})\left(
b_{1}-b_{2}\right) }{\left\Vert \mathcal{Z}_{s}(\rho _{d\times d})\left(
b_{1}-b_{2}\right) \right\Vert _{\mathbb{R}^{3}}}\in \mathbb{R}^{3},  \notag
\end{eqnarray}%
and is equal to%
\begin{equation}
\left\Vert \mathcal{Z}_{s}(\rho _{d\times d})\left( b_{1}+b_{2}\right)
\right\Vert _{\mathbb{R}^{3}}+\left\Vert \mathcal{Z}_{s}(\rho _{d\times
d})\left( b_{1}-b_{2}\right) \right\Vert _{\mathbb{R}^{3}}.  \label{15}
\end{equation}%
Therefore,%
\begin{align}
 &\max_{\substack{ a_{m},b_{k}\in \mathbb{R}^{3}, \\ \left\Vert
a_{m}\right\Vert ,\left\Vert b_{k}\right\Vert =1}}\left\vert \left\langle 
\mathcal{B}_{chsh}(a_{1},a_{2};b_{1},b_{2})\right\rangle _{\rho _{d\times
d}}\right\vert    \label{16_} \\= &\max\left\lbrace
\left\Vert \mathcal{Z}_{s}(\rho _{d\times d})\left( b_{1}+b_{2}\right)
\right\Vert _{\mathbb{R}^{3}}+\left \Vert \mathcal{Z}_{s}(\rho _{d\times
d})\left( b_{1}-b_{2}\right) \right\Vert _{\mathbb{R}^{3}}\right\rbrace ,   \nonumber
\end{align}
where the maximum in the second line is taken over the set of all unit vectors $b_{k}\in \mathbb{R}^{3}$. 

Taking further into account that
\begin{eqnarray}
\left\Vert b_{1}\pm b_{2}\right\Vert^{2}_{\mathbb{R}^{3}} &=&2\pm 2(b_{1},b_{2}) _{\mathbb{R}^{3}},\text{ \ }
\label{17} \\
(b_{1},b_{2})_{\mathbb{R}^{3}} &=&\cos \theta ,\text{ \ }\theta \in \lbrack 0,\pi ],  \notag
\end{eqnarray}%
let us introduce (like in \cite{Hor.Hor.Hor:95}) two unit mutually
orthogonal vectors $r_{1},r_{2}\in \mathbb{R}^{3}$, satisfying the relations 
\begin{align}
\frac{b_{1}+b_{2}}{2} =r_{1}\cos \theta ^{\prime },&\text{\ \ \ \ }\frac{%
b_{1}-b_{2}}{2}=r_{2}\sin \theta ^{\prime },\ \ \theta ^{\prime }=\frac{\text{\ }\theta }{2}\in \lbrack 0,\frac{%
\pi }{2}], \label{18}\\
 2\left( r_{1},r_{2}\right)_{\mathbb{R}^{3}} \sin \theta & =\left\Vert b_{1}\right\Vert
^{2} _{\mathbb{R}^{3}}-\left\Vert b_{2}\right\Vert^{2} _{\mathbb{R}^{3}}=0.  \notag
\end{align}%
Due to (\ref{18}), we come in (\ref{16_}) to the following expression for the maximum in (\ref{13}): 
\begin{align}
& \max_{\substack{ a_{m},b_{k}\in \mathbb{R}^{3}, \\ \left\Vert
a_{m}\right\Vert ,\left\Vert b_{k}\right\Vert =1}}\left\vert \left\langle 
\mathcal{B}_{chsh}(a_{1},a_{2};b_{1},b_{2})\right\rangle _{\rho _{d\times
d}}\right\vert   \label{16b_} \\
& =2\max \left\lbrace \cos \theta^{\prime}\left\Vert \mathcal{Z}_{s}(\rho _{d\times d})\left(
r_{1}\right) \right\Vert _{\mathbb{R}^{3}}+\sin \theta^{\prime}\left\Vert \mathcal{Z}%
_{s}(\rho _{d\times d})\left( r_{2}\right) \right\Vert _{\mathbb{R}%
^{3}}\right\rbrace ,  \notag
\end{align}%
where, in the second line, the maximum is taken over all angles $\theta
^{\prime }\in \lbrack 0,\frac{\pi }{2}]$ and all mutually orthogonal unit
vectors $r_{1},r_{2}\in \mathbb{R}^{3}.$ 

Clearly, the maximum in the second line of (\ref{16b_}) over angle 
$\theta^{\prime}$ is attained at the angle $\theta_{0}^{\prime}$, defined by the relation 
\begin{equation}
\text{tg}\theta_{0} ^{\prime }=\frac{\left\Vert \mathcal{Z}_{s}(\rho
_{d\times d})\text{ }r_{2}\right\Vert }{\left\Vert \mathcal{Z}_{s}(\rho
_{d\times d})\text{ }r_{1}\right\Vert} 
\end{equation}
and is equal to 
\begin{equation}
\sqrt{\left\Vert \mathcal{Z}_{s}(\rho _{d\times d})r_{1}\right\Vert _{%
\mathbb{R}^{3}}^{2}+\left\Vert \mathcal{Z}_{s}(\rho _{d\times
d})r_{2}\right\Vert _{\mathbb{R}^{3}}^{2}}.  \label{20}
\end{equation}%
Noting that the square of the norm  
\begin{equation}
\left\Vert \mathcal{Z}_{s}(\rho _{d\times d})r_{j}\right\Vert _{\mathbb{R}%
^{3}}^{2}=\left( r_{j},\mathcal{Z}_{s}^{T}(\rho _{d\times d})\mathcal{Z}%
_{s}(\rho _{d\times d})r_{j}\right) _{\mathbb{R}^{3}} \label{59}
\end{equation} 
and the local maxima of 
the scalar product standing in the right-hand side of (\ref{59}) are attained at the mutually orthogonal eigenvectors 
 of the positive operator $\mathcal{Z}_{s}^{T}(\rho _{d\times d})\mathcal{Z%
}_{s}(\rho _{d\times d}),$ 
we derive that the maximum of the radical expression in (\ref{20}) over all mutually orthogonal vectors $r_{1},r_{2}\in \mathbb{R}^{3}$ is attained at two eigenvectors of $\mathcal{Z}_{s}^{T}(\rho _{d\times d})\mathcal{Z%
}_{s}(\rho _{d\times d}),$ corresponding to the largest eigenvalues of this operator, and is equal to the sum of these two largest eigenvalues, that is, the sum of the squares of two largest singular values of matrix $\mathcal{Z}_{s}(\rho _{d\times d})$. This proves the statement of Theorem \ref{Theorem1}. 
\end{proof}

\section*{Appendix B}
For the pure state (\ref{y}), the coefficients $\zeta_{m m^{\prime}, kk^{\prime}}$ in decomposition (\ref{22}) are equal to $\sqrt{\mu_{m}\mu_{m}^{\prime
}}\delta_{mk}\delta_{m^{\prime}k^{\prime}}$, so that by (\ref{Z11d})--(\ref{Z31d}) the non-diagonal elements of matrix $\mathcal{Z}_{s}
(|\psi_{d\times d}\rangle )$ are equal to zero while the diagonal elements are:
\begin{align}
\mathcal{Z}_{s}^{(11)}
(|\psi_{d\times d}\rangle )    &=\frac{1}{2}\sum_{m,k=1}%
^{d-1}\sqrt{mk(d-m)(d-k)\mu_m \mu_{m+1}}\delta_{mk},\label{g1}\\
&=\frac{1}{2}\sum_{k=1}^{d-1}k(d-k)\sqrt{\mu _{k}\mu _{k+1}}=\sum_{k=1}^{2s}k\left(s-\frac{k-1}{2}\right)\sqrt{\mu _{k}\mu _{k+1}},\nonumber\\
-\mathcal{Z}_{s}^{(22)}
(|\psi_{d\times d}\rangle )     &=\frac{1}{2}\sum_{m,k=1}%
^{d-1}\sqrt{mk(d-m)(d-k)\mu_m \mu_{m+1}}\delta_{mk} ,\nonumber\\
&=\frac{1}{2}\sum_{k=1}^{d-1}k(d-k)\sqrt{\mu _{k}\mu _{k+1}}=\sum_{k=1}^{2s}k\left(s-\frac{k-1}{2}\right)\sqrt{\mu _{k}\mu _{k+1}},\nonumber\\
\mathcal{Z}_{s}^{(33)}
(|\psi_{d\times d}\rangle )     &=\frac{1}{4}\sum_{m,k=1}%
^{d}(d+1-2m)(d+1-2k)\mu_m \delta_{mk} \nonumber\\
& =\frac{1}{4}\sum_{k=1}%
^{d}(d+1-2k)^2 \mu_k=\sum_{k=1}^{2s+1}\left( s-(k-1)\right) ^{2}\mu _{k}.\nonumber 
\end{align}

\section*{Appendix C}
From relations (\ref{coeff_werner}) and (\ref{Z11d})--(\ref{Z31d})
it follows:
\begin{align}
\mathcal{Z}^{(12)}_{s}(\rho_{d,\Phi}^{(wer)})=\mathcal{Z}^{(21)}_{s}(\rho_{d,\Phi}
^{(wer)})=\mathcal{Z}^{(23)}_{s}(\rho_{d,\Phi}^{(wer)}) =\mathcal{Z}^{(32)}_{s}
(\rho_{d,\Phi}^{(wer)})=0\label{g2}
\end{align}
and
\begin{align}
\mathcal{Z}_{s}^{(11)}(\rho_{d,\Phi}^{(wer)}) &  =\frac{1}{2}\sum
_{m,k=1}^{d-1}\text{ }\sqrt{mk(d-m)(d-k)}\ \frac{d\Phi-1}{d(d^{2}-1)}%
\delta_{mk}\delta_{(m+1)(k+1)}=\frac{d\Phi-1}{12},\label{g3}\\
\mathcal{Z}_{s}^{(22)}(\rho_{d,\Phi}^{(wer)}) &  =\frac{1}{2}\sum
_{m,k=1}^{d-1}\text{ }\sqrt{mk(d-m)(d-k)}\ \frac{d\Phi-1}{d(d^{2}-1)}%
\delta_{mk}\delta_{(m+1)(k+1)}=\frac{d\Phi-1}{12},\nonumber\\
\mathcal{Z}_{s}^{(33)}(\rho_{d,\Phi}^{(wer)}) &  =\frac{1}{4}\sum_{m,k=1}%
^{d}(d+1-2m)(d+1-2k)\left(  \frac{d-\Phi}{d(d^{2}-1)}+\frac{d\Phi-1}%
{d(d^{2}-1)}\delta_{mk}\right) \nonumber\\
&  =\frac{1}{4}\frac{d\Phi-1}{d(d^{2}-1)}\sum_{m=1}^{d}(d+1-2m)^{2}=\frac
{1}{4}\frac{d\Phi-1}{d(d^{2}-1)}\frac{1}{3}d(d^{2}-1)=\frac{d\Phi-1}%
{12}.\nonumber
\end{align}
Also,
\begin{align}
\mathcal{Z}^{(31)}_{s}(\rho_{d,\Phi}^{(wer)}) &=\mathcal{Z}^{(13)}_{s}(\rho_{d,\Phi}^{(wer)})
\label{g4}\\
&  =\frac{1}{2}\sum_{m,k=1}^{d}\text{
}\sqrt{m(d-m)}\left(  d+1-2k\right)  \left(  \frac{d-\Phi}{d(d^{2}-1)}%
+\frac{d\Phi-1}{d(d^{2}-1)}\right)  \delta_{(k+1)k}=0. \nonumber
\end{align}

\end{document}